\documentclass[11pt,a4paper]{article}
\pdfoutput=1

\usepackage{mathtools}
\usepackage{mathrsfs}
\usepackage{amsmath}
\usepackage{tikz-feynman}
\usepackage[utf8x]{inputenc}
\usepackage{amssymb}
\usepackage{slashed}
\usepackage{bm}
\usepackage{color}
\usepackage{tikz}
\usepackage{cite}
\usepackage{braket}
\usepackage{float}
\usepackage[]{todonotes}
\usepackage{jheppub}
\usepackage{twistor}

\renewcommand{\[}{\begin{equation}\begin{aligned}}
\renewcommand{\]}{\end{aligned}\end{equation}}

\newcommand{\ud}{{\mathrm{d}}}

\title{Classical physics from amplitudes on curved backgrounds}
\author[1]{Tim Adamo,}
\author[1]{Andrea Cristofoli,}
\author[2]{Anton Ilderton}

\affiliation[1]{School of Mathematics and Maxwell Institute for Mathematical Sciences\\
University of Edinburgh, EH9 3FD, United Kingdom}
\affiliation[2]{Higgs Centre, School of Physics and Astronomy\\ University of Edinburgh, EH9 3FD, United Kingdom}

\emailAdd{t.adamo@ed.ac.uk}
\emailAdd{acristof@exseed.ed.ac.uk}
\emailAdd{anton.ilderton@ed.ac.uk}

\abstract{We generalise the Kosower-Maybee-O'Connell (KMOC) formalism relating classical observables and scattering amplitudes to curved backgrounds. We show how to compute the final semiclassical state for a particle moving in a curved background in terms of scattering amplitudes on that background. Two-point amplitudes in this framework correspond to conservative physics with background-dependent memory effects. As an application, we consider plane wave and shock wave backgrounds both in electromagnetism and general relativity. We determine the final semiclassical state, showing it satisfies a notion of double copy on curved backgrounds. We then conclude by computing the impulse of a particle on such backgrounds, deriving exact results and velocity memory effects.
}

\begin{document}
\maketitle
\section{Introduction}
The Kosower-Maybee-O'Connell (KMOC) formalism is a successful framework in which to study how classical physics arises from on-shell scattering amplitudes on flat space-time \cite{Kosower:2018adc}. Over the past few years, it has demonstrated remarkable and deep relations between classical observables and the classical limit of scattering amplitudes. For instance, $6$-point amplitudes have been used to study the zero variance properties of observables in the classical limit \cite{Cristofoli:2021jas,Britto:2021pud}, while $5$-point amplitudes and soft emission encode leading-order waveforms and the radiative Newman-Penrose scalar in scattering processes \cite{Cristofoli:2021vyo,Bautista:2021inx,Bautista:2021llr,Bautista:2019tdr}. Similarly, tree-level $4$-point amplitudes have been used to study conservative physics such as the impulse in general relativity, Yang-Mills and modified gravity \cite{Maybee:2019jus,Guevara:2019fsj,Aoude:2021oqj,delaCruz:2020bbn,delaCruz:2021gjp,Burger:2019wkq}, and cuts of two-loop, $4$-point amplitudes relate to radiative physics such as the total radiated momentum in classical scattering \cite{Herrmann:2021lqe,Herrmann:2021tct}. More recently, $3$-point amplitudes for massive spinning particles have been used to elucidate the Newman-Janis shift \cite{Arkani-Hamed:2019ymq}, while in split signature it has been shown that they correspond to the Maxwell and Weyl curvature spinors of linear solutions \cite{Monteiro:2020plf,Monteiro:2021ztt}.

Considering these successes, one is tempted to ask if lower-point amplitudes could play any role in this setup, but the answer would seem to be negative: $2$-point amplitudes are trivial on a flat space-time. However, when computed on curved backgrounds, such amplitudes can reveal an incredibly rich structure such as memory effects, eikonal behaviour and even stringy factorization properties (cf., \cite{Gibbons:1975jb,tHooft,Garriga:1990dp,Verlinde:1991iu,Jackiw:1991ck,Kabat:1992tb,Adamo:2017nia,Adamo:2021jxz,Adamo:2021hno,Adamo:2021rfq}). A natural question is then: is it possible to generalise the KMOC formalism to extract classical physics from amplitudes, at 2-points and beyond, on a curved background? It seems natural to expect that such a generalization would provide insight into the classical limit of non-perturbative observables that are typically studied in strong field QED~\cite{Fedotov:2022ely} and QFT on curved spacetime \cite{DeWitt:1975ys}.


Besides broadening the range of observables accessible
to KMOC~\cite{Kosower:2018adc}, these questions also
relate to the fact that when considering the dynamics of classical particles, we have implicitly fixed a background in which those particles are moving. Viewing KMOC as essentially
a framework for extracting classical information from quantum field theory (QFT), this
arises from fixing a background around which to perform the perturbative expansion of
QFT. However, at the level of first principles there is nothing (besides computational simplicity) which requires that the background be trivial: indeed, any classical solution to
the equations of motion can be used as a background for perturbation theory in the background field formulation of QFT (cf., \cite{Furry:1951zz,DeWitt:1967ub,tHooft:1975uxh,Abbott:1981ke}). Using a curved background in QFT should then translate, in the classical limit, into classical particle dynamics in a curved background: we will see that this is indeed the case, thus generalizing the main idea presented in \cite{Kosower:2018adc} beyond a trivial background.


\medskip

This paper is organized as follows. In section~\ref{sec:CB}, we generalise the KMOC formalism by considering QFT on a curved background, assumed to be an asymptotically flat solution of the Maxwell or Einstein equations which admits an S-matrix. We show how to express the final semiclassical state in terms of $n$-point amplitudes on a curved background, focusing for the rest of the paper on the contributions arising from $2$-points. In sections~\ref{sec:PW} and \ref{sec:SW} we consider the examples of (vacuum) plane wave and shockwave solutions, respectively; in both cases we look at electromagnetism and gravity. We give expressions for the final semiclassical state describing a single point particle crossing such backgrounds -- in the same spirit of \cite{Cristofoli:2021jas} -- and provide a notion of double copy for final states. We will show in particular that the final semiclassical state on a plane wave or shockwave background assumes a universal form where all the classical information carried by the background is dressed in a shifted momentum of the wavepacket, both in electromagnetism and gravity. Finally, we extract the physical observable of the impulse (without radiation) from the final semiclassical state. For shockwaves, we find agreement with the geodesics on such backgrounds (e.g., \cite{tHooft,Klimcik:1988az}). For plane waves, we recover the `velocity memory effect' for a point particle crossing a plane wave (e.g., \cite{Zhang:2017rno,Zhang:2018srn,Shore:2018kmt,Steinbauer:2018iis,Bieri:2020zki,DiVecchia:2022owy}), which so far has not been derived in KMOC on a flat space-time.

\section{KMOC on curved backgrounds}
\label{sec:CB}

Consider the classical dynamics of a point particle following the geodesics on an exact, asymptotically flat solution to the Einstein field equations, or the classical dynamics under the Lorentz force of a charged point particle in an exact, asymptotically flat solution to the Maxwell equations. Our aim is to develop a formalism, analogous to KMOC~\cite{Kosower:2018adc}, relating the observables of this classical system to on-shell scattering amplitudes in a curved background; as such, we assume that the background solutions admit an S-matrix\footnote{In the sense that there is unitary evolution between the asymptotically flat regions and no spontaneous pair production in the free theory~\cite{Hawking:1975vcx,Gibbons:1975jb,Gibbons:1975kk,Woodhouse:1976fe}.}. To this end, we start by considering the initial state of the system in the asymptotically flat region
\begin{equation}\label{ini}
    \ket{\Psi}=\int \d\Phi(p)\, \phi_{b}(p)\,  \ket{p} \ ,
\end{equation}
where $\phi_{b}(p):=\phi(p)\, \e^{\im p\cdot b/\hbar}$, $\phi(p)$ is a properly normalized wavepacket, $b^{\mu}$ is a four-vector encoding information on the origin of the reference frame, and
\begin{equation}\label{OSmeasure}
   \ud \Phi(p) :=
   \frac{\d^4p}{(2 \pi)^4}
   \: 2 \pi\, \Theta\!\left(p^{0}\right)\, \delta\!\left(p^{2}-m^{2}\right)\, .
\end{equation}
Within this region of space-time, the wavepacket $\phi(p)$ is chosen to provide a well-defined notion of classical particle dynamics. More precisely, its Compton wavelength $\ell_c$, wavepacket spread $\ell_w$ and scattering length $b^{\mu}$ obey the `Goldilocks' relations $\ell_c\ll\ell_w\ll\sqrt{-b^2}$~\cite{Kosower:2018adc}.

We denote the $S$-matrix on the electromagnetic or gravitational background by $\mathcal{S}$; since the background is treated exactly and without approximation (i.e., non-perturbatively), $\mathcal{S}$ itself depends on and contains the background fields. Interactions between the initial state \eqref{ini} and the exact background can be expressed in terms of the final state
\begin{equation} \label{deffin}
\mathcal{S}\ket{\Psi}=\int \d\Phi(p)\, \phi_{b}(p)\, \mathcal{S}\ket{p} \ .
\end{equation}
Following~\cite{Cristofoli:2021jas}, we introduce a completeness relation in the integrand of \eqref{deffin}, so that the final state is fixed in terms of $n$-point scattering amplitudes on the curved background\footnote{We use a shorthand notation for the integration measure where $\d\Phi(p_1,..p_n):=\prod_{j=1}^{n}\d\Phi(p_{j})$.}
\begin{equation}\label{npoint}
    \mathcal{S}\ket{\Psi}=\sum_{\mathrm{states}}\int \d\Phi(p,p',k_1,\ldots,k_n)\: \phi_{b}(p) \:
    \sum_{
    \eta_1,..\eta_n}\ket{p',k_1^{\eta_1},\ldots,k_n^{\eta_n}}\bra{p',k_1^{\eta_1},\ldots,k_n^{\eta_n}}\mathcal{S}\ket{p} \;.
\end{equation}
Here, the first sum runs over all massless states that can be emitted during the scattering process (i.e., photons or gravitons), while the second sum is over the quantum numbers $\eta$ (e.g., polarizations) of each emitted state.

Neglecting the emission of radiation, we have
\begin{align} \label{one}
     \mathcal{S}\ket{\Psi}=\int \d\Phi(p,p')\: \phi(p) \:\e^{\im\, p 
     \cdot b/\hbar}\,
    \bra{p'}\mathcal{S}\ket{p}\,\ket{p'}\,,
\end{align}
which implies that the conservative dynamics of a point particle on a curved background is fully captured by the $2$-point amplitude on the background. While these 2-point amplitudes vanish in trivial backgrounds, they are generically non-zero in curved backgrounds. To provide some concrete examples, we now evaluate the final semiclassical state \eqref{one} on exact plane wave and shockwave backgrounds. This amounts to the evaluation of the associated $2$-point amplitude, which we consider both in electromagnetism and in general relativity. 

\section{Plane wave backgrounds}
\label{sec:PW}
Vacuum plane waves are highly-symmetric solutions of electromagnetism and gravity which have a natural physical interpretation as coherent states of photons or gravitons. Furthermore, asymptotically flat `sandwich' plane waves admit an S-matrix~\cite{Schwinger:1951nm,Gibbons:1975jb,Garriga:1990dp,Adamo:2017nia} as well as WKB-exact solutions to background coupled wave equations~\cite{Wolkow:1935zz,Ward:1987ws,Mason:1989,Adamo:2017nia}.


\subsection{Final states and double copy}

We start by considering a vacuum plane wave solution to Maxwell equations. It is convenient to work in lightfront coordinates for which the line element is
\be\label{ds-lf} 
    \mathrm{d} s^{2}=2 \mathrm{~d}x^{+}  \mathrm{d}x^{-}-\mathrm{d}x^{\perp} \mathrm{d}x^{\perp} \ ,
\ee
using $\perp$ as a shorthand notation for $b=1,2$. In these coordinates, plane wave solutions are given by electromagnetic 4-potentials of the form
\begin{equation}\label{yildiz}
    \mathrm{A}_{\mu}(x)= 
    -x^{\perp}\, {E}_{\perp}(x^-)\, n_{\mu}  \ , 
\end{equation}
where $n_{\mu}=\delta_{\mu}^{-}$ and ${E}_{\perp}(x^-)$ are the $2$ (transverse) electric field components, depending on the lightfront variable $x^{-}$. We take the electric fields to be `sandwich': supported on a finite range $x_i^- < x^- < x_f^-$ but otherwise of arbitrary profile. 
We will also need the quantity
\begin{equation}\label{a-int-E}
    a_{\perp}(x^{-}) :=\int_{-\infty}^{x^{-}} \d s \: E_{\perp}(s)\,,
\end{equation}
which is the `work done' on a charged particle as it crosses the plane wave, entering from the asymptotic past.

%
The scalar 2-point amplitude is computed by a boundary term in the classical action (see Section 2 of~\cite{Adamo:2021rfq} for a recent review); from~\cite{Adamo:2017nia}, the tree-level charged scalar $2$-point amplitude on an electromagnetic plane wave background is, in terms of \eqref{a-int-E},
\begin{equation}\label{taban}
\bra{p'}\mathcal{S}\ket{p}= 2p_{+}\,\hat{\delta}\!\left(p'_+ - p_+\right)\,\hat{\delta}^2\!\left(p'_\perp- p_\perp+ea_{\perp}(\infty)\right)\, \e^{\im\,\tilde{s}_{p}/\hbar} \;,
\end{equation}
where $\hat{\delta}(x):=2\pi \delta(x)$, $e$ is the charge of the scalar particle (incoming and outgoing, due to charge conservation) and 
\begin{equation}
    \tilde{s}_p:=\int_{x_i^-}^{x^-_f}\d s\,s \frac{\ud}{\ud s} \frac{e^2a^2_\perp(s) - 2ep_\perp a_\perp(s)}{2\,p_+}\,.
\end{equation}
The quantity $a_{\perp}(\infty)$ is related to the zero-frequency modes of the electric field~\cite{Dinu:2012tj},
\begin{equation}\label{QEDzeromode}
    a_{\perp}(\infty):=\int_{-\infty}^{+\infty} \d x^{-} \: E_{\perp}(x^{-})=\tilde{E}_{\perp}(0)\,,
\end{equation}
where $\tilde{E}(\omega)$ denotes the Fourier transform.
Note that this can be non-zero even though the physical fields vanish outside the sandwich region. The implication, given that $a_\perp$ is in some sense the work done, is that there can be a net change in momentum delivered to a classical particle crossing a plane wave; this is of course an example of the electromagnetic (velocity) memory effect~\cite{Dinu:2012tj,Bieri:2013hqa,Susskind:2015hpa,Pasterski:2015zua}, as we will investigate below. 


The final state (\ref{one}) can then be expressed in the compact but explicit form
\begin{equation}\label{memo}
    \mathcal{S}\ket{\Psi}=\int \ud\Phi(p')\: e^{\mathrm{i} \tilde{s}_{l}/\hbar}\,
    \phi_{b}(p^{\prime}_+,p'_\perp +ea_{\perp}(\infty))\,
    \ket{p'} \quad 
 \ ,
\end{equation}
in which we have highlighted that, because of the on-shell condition, the wavepacket $\phi_b(p)$ is a function of only three independent momentum variables, which it is natural to choose to be $p_+$ and $p_\perp$.

\medskip

We turn now to the analogue final state on a \textit{gravitational} plane wave. These are vacuum solutions to the Einstein equations which can be represented, in Brinkmann coordinates, as
\begin{equation}\label{altingun}
\mathrm{d} s^{2}=2 \mathrm{~d} x^{+} \mathrm{~d} x^{-}- H_{a b}(x^{-})\, x^{a} x^{b}\,(\mathrm{d}x^{-})^2-\mathrm{d}x^{\perp} \mathrm{~d}x^{\perp} \;,
\end{equation}
in which $H_{ab}(x^-)$ is a trace-free $2\times2$ matrix with support on a compact `sandwich' region $[x^-_{i},x^-_{f}]$ but otherwise arbitrary. 

Of particular relevance in the study of gravitational memory effects \cite{Strominger:2017zoo,Shore:2018kmt} are the transverse zweibeins which satisfy the geodesic deviation equations $\ddot{E}_{a i}=H_{a b} E_{i}^{b}$. Just like the electromagnetic case, constants of integration for these equations are fixed by boundary conditions; we impose
\begin{equation}\label{grbc}
E^{a}_{i}(x^-<x_i^-)=\delta^{a}_{i}\,, \qquad E^{a}_{i}(x^->x_f^-)=\delta^{a}_i+x^-\,c^{a}_{i}\,,
\end{equation}
where $c^{a}_i$ is a non-vanishing constant matrix fixed by the boundary conditions and $H_{ab}(x^-)$; this is the gravitational analogy of $a_\perp(\infty)$. It is easy to see that despite linear $x^-$-dependence, $E^{a}_i$ does indeed describe flat Minkowski space in the $x^->x^-_f$ region.

In terms of these quantities, the $2$-point amplitude is~\cite{Garriga:1990dp,Adamo:2017nia}
\vspace{1mm}
\begin{equation}\label{sincap}
\left\langle p^{\prime}|\mathcal{S}| p\right\rangle=2 p_{+}\,  \hat{\delta}\!\left(p_{+}^{\prime}-p_{+}\right)\, F\!\left(\mathrm{p}^{\prime}_{\perp}-\mathrm{p}_{\perp}, p_{+}\right) \quad , \end{equation}
\begin{equation}\label{eq:gau-f}
    F\!\left(\mathrm{q}_{\perp}, p_{+}\right):=\frac{2 \pi}{p_{+} \: \sqrt{\left| \operatorname{det}(c)\right|}\: \hbar }\, \exp\! \left(-\frac{\im}{2 p_{+}\, \hbar } \:  \mathrm{q}_{\perp} \cdot c^{-1} \cdot \mathrm{q}_{\perp}\right) \ ,
\end{equation}
\newline
where $c^{a}_{b}$ is shorthand for $c^{a}_{i}\,\delta^{i}_{b}$. This yields the following expression for the final semiclassical state \eqref{one} in gravity
\begin{equation}\label{GRfinalPW}
\mathcal{S}|\Psi\rangle=\int \ud \Phi(p, p^{\prime})\,2\,p_+\, \phi_{b}(p) \, \hat{\delta}\!\left(p_{+}^{\prime}-p_{+}\right)\, F\!\left(\mathrm{p}^{\prime}_{\perp}-\mathrm{p}_{\perp}, p_{+}\right)\, \left|p^{\prime}\right\rangle \,.
\end{equation}
This expression can be further simplified by changing variable $p\to q =p-p'$, which gives
\begin{equation}
\mathcal{S}|\Psi\rangle=\int \ud \Phi(p^{\prime})\,2\,p_+^{\prime}\, e^{ip' \cdot b/\hbar} \int \hat{d}^4q\, \hat{\delta}\!\left(q_{+}\right) \hat{\delta}(2q \cdot p')\: \phi(p'+q)
e^{i q\cdot b/ \hbar}F\!\left(\mathrm{q}_{\perp}, p_{+}\right)\, \left|p^{\prime}\right\rangle \,,
\end{equation}
the benefit being that the integrals over $q_{+}$ and $q_{-}$ become straightforward. Following~\cite{Cristofoli:2021jas}, we can perform the remaining integral over $q_{\perp}$ via a stationary phase approximation -- since we are considering our expressions in the limit of $\hbar \rightarrow 0$, this approximation 
effectively becomes exact.
The stationary point $q^{*}_{\perp}$ is given by
\begin{equation}
    q^{*}_{a}= p_{+}\: c_{a}^{i} b_{i} \ ,
\end{equation}
which yields an expression for the final semiclassical state in gravity that is remarkably close to the electromagnetic case (\ref{memo}),
\begin{equation}\label{eq:state-in-gr-final}
    \mathcal{S}|\Psi\rangle=\int \ud \Phi(p^{\prime})\,e^{i\delta(q^{*})/\hbar} \phi(p'_+,p^{\prime}_{\perp}+p_{+}^{\prime}\: c_{\perp}^{i} b_{i})\, \left|p^{\prime}\right\rangle \,.
\end{equation}
where $\delta(q^{*})$ is an $\hbar$ independent phase which arises after stationary phase.

\medskip

\paragraph{Double copy for final states:} Before proceeding to analyse the physics contained in the 2-point amplitudes and associated final states, we point out that the final semiclassical state in electromagnetism and gravity can be understood from the point of view of the double copy on plane wave backgrounds introduced in~\cite{Adamo:2017nia}.
The starting point is given by the amplitudes (\ref{taban}) and (\ref{sincap}) on electromagnetic and gravitational plane wave backgrounds, respectively. The most obvious difference between the two is the number of momentum-conserving delta functions; this is despite the waves having the same symmetry group. Another puzzle is what to do with the asymptotic gauge field appearing inside the delta functions in the electromagnetic case. These two points are related; to see how, we must first undo the integral over the 2-dimensional transverse coordinates which was part of the 2-point amplitude, writing the delta function as
\be\label{QEDfas}
   \hat{\delta}^2\!\left(p'_\perp- p_\perp+ea_{\perp}(\infty)\right)= \int\!\ud^2 x^{\perp}\, \exp\left[\mathrm{i}\, x^\perp\,( p'-p + e\, a(\infty))_\perp\right] \,.
\ee
The double copy rules of~\cite{Adamo:2017nia} can now be applied, in particular the replacement
\begin{equation}
    \left.e\, a^i(x^-)\right|_{x^->x^-_f} \longrightarrow \frac{p_+}{2}\, c_{a}^{i}\,x^{a} \;,
\end{equation}
which is motivated by the observation that the non-trivial component of the electromagnetic plane wave is a linear function of $x^{a}$ \eqref{yildiz} while the non-trivial component of a gravitational plane wave is quadratic \eqref{altingun}. 

As a consequence of this double copy procedure, we obtain in the gravitational amplitude
\be
  \int\!\ud^2 x^{\perp}\, \exp\left[\mathrm{i}\, x^\perp\,(p'-p)_\perp + \im\, p_+\, \frac{c_{ab}}{2}\, x^a x^b\right] \; .
\ee
It is easy to see that the resulting Gaussian integral yields the desired result \eqref{sincap} for gravity, thus providing a double copy explanation for the structure of this amplitude.

\subsection{Memory effects from amplitudes on plane wave backgrounds}

Equipped with the knowledge of the conservative final state in both the electromagnetic and gravitational cases, we can proceed to compute observables. A key example for conservative physics is the change in momentum -- or \emph{impulse} -- experienced by a particle crossing a plane wave background
\begin{equation}\label{general}
    \Delta p^{\mu}:= \bra{\Psi}\mathcal{S}^{\dag}\mathbb{P}^{\mu}\mathcal{S}\ket{\Psi}-\bra{\Psi}\mathbb{P}^{\mu}\ket{\Psi}\,,
\end{equation}
where $\mathbb{P}_\mu$ is the momentum operator. Let us start with the electromagnetic case. Using the final state \eqref{memo}, the impulse becomes
\begin{equation}
  \Delta p^{\mu}=\int \ud\Phi(p')\, \left|\phi(p'_+,\,p'_\perp+e\,a_\perp(\infty))\right|^2\,p^{\prime\mu} -p^{\mu} \,,
\end{equation}
where $p^{\mu}$ is the incoming momentum. Taking the classical limit, one obtains
\begin{equation}\label{eq:QEDresult}
    \Delta p^{\mu}= -e\,a^{\mu}(\infty)+n^{\mu}\bigg(\frac{2e\: p\cdot a(\infty)- e^2\,a^2(\infty)}{2\,p_+}\bigg)\,\ ,
\end{equation}
where $a_\mu=(0,0,a_\perp)$. 

This result is in agreement with the literature~\cite{Taub:1948zza,Sengupta}, and has previously been derived, including $\mathcal{O}(e^2)$ radiative corrections, from the classical limit of QED amplitudes in plane wave backgrounds~\cite{Ilderton:2013dba,Ilderton:2013tb}. For reference, a purely classical derivation is given in Appendix~\ref{AppA}. 
The same observable has also been studied in the flat background formulation of KMOC~\cite{Cristofoli:2021vyo} by considering the interactions between a coherent state and a semiclassical state describing a classical point particle. Interestingly, it was found that the leading order non-vanishing contribution to the impulse started at $e^2$, in contrast with the linear contribution appearing in \eqref{eq:QEDresult}. In our framework, the non-vanishing impulse of the particle at order $e$ is related to the non-zero net work done \eqref{QEDzeromode} by the electric field and it is known as the electromagnetic velocity memory effect \cite{Campoleoni:2019ptc,Pasterski:2015zua,Susskind:2015hpa}.  From the perspective of scattering amplitudes on a flat background \cite{Cristofoli:2021vyo}, one can derive this result by noticing that the `in' and `out' vacua are related by a large gauge transformation, which must be incorporated into the amplitude by modifications of the LSZ formulae~\cite{Kibble:1965zza,Dinu:2012tj}\footnote{Similar asymptotic dressings by large gauge transformations arise in the study of IR divergenes in QED; see~\cite{Mirbabayi:2016axw,Gabai:2016kuf,Kapec:2017tkm} for recent investigations.}. These issues are avoided here by treating the background non-perturbatively\footnote{For alternative ways to obtain such memory effects, see also \cite{Manohar:2022dea,DiVecchia:2022owy}.}. 

\medskip

Let us now turn our attention to gravitational plane waves. Using the final state derived in (\ref{eq:state-in-gr-final}) we have that the impulse experienced by a particle crossing a gravitational plane wave background - in absence of emitted radiation - is
\begin{equation}
 \Delta p^{\mu}=\int \mathrm{d} \Phi\left(p^{\prime}\right)\left|\phi\left(p_{+}^{\prime}, p_{\perp}^{\prime}+p_{+} c_{\perp}^{i} b_{i}\right)\right|^{2} p^{\prime \mu}-p^{\mu}   
\end{equation}
A straightforward calculation shows that the final result is
\begin{equation}\label{no}
    \Delta p^{\mu}= p_{+}\,\delta^{\mu}_{a}\,c^{a}_{i}\,b^{i}-n^{\mu}\, \bigg(\frac{2 p_{+}\: p_{a}c^{a}_{i}b^{i}-p^2_{+}c^a_{i}b^{i}c^{b}_{j}b^{j}\delta_{ab}}{2p_{+}} \bigg) \,,
\end{equation}
where we recall from \eqref{grbc} that $c^{a}_i$ corresponds to $\dot{E}^{a}_i$ for the transverse zweibein, evaluated in the far future, $x^->x_f^-$.

Removing the masses from both sides of this relation, this result is equivalent to the change in the geodesic velocity along the $\perp$ direction
\begin{equation}
    \Delta u^{a}= \frac{p_{+}}{m}\, \left.\dot{E}^{a}_{j}\right|_{x^->x_f^-}\,b^{j}\,,
\end{equation}
in agreement with the so called velocity memory effect\footnote{We refer to Appendix~\ref{AppA} for a classical derivation of this effect.} in general relativity (e.g., \cite{Shore:2018kmt}). The dependence on the origin of the frame $b^{\mu}$, absent in the electromagnetic case, is a consequence of the lack of translation invariance in a gravitational plane wave. 


\section{Shock wave backgrounds}
\label{sec:SW}

We now turn our attention to shock wave backgrounds. In the electromagnetic case, shock waves represent exact solutions of the Maxwell equations sourced by a massless charged particle. In lightfront coordinates \eqref{ds-lf}, the shock wave potential can be written as~\cite{Jackiw:1991ck}
\begin{equation}
    \mathrm{A}_{\mu}(x)=-\frac{e}{2 \pi}\,\delta(x^-)\, n_{\mu}\,\log(|x^{\perp}|) \, ,
\end{equation}
where $e$ is the charge of the shock wave background and, as before, $n_{\mu}=\delta^{-}_{\mu}$. Physically, these fields can be understood as arising from the ultra-relativistic boost of a Coulomb field. The natural analogue in gravity is the Aichelburg-Sexl metric describing the gravitational field of an ultra-boosted Schwarzschild black hole \cite{Aichelburg,Dray:1984ha}:
\begin{equation}
    \ud s^2= 2\, \ud x^{+}\,\ud x^{-}- \ud x^{\perp} \ud x^{\perp}+8G_{N}\,P_{-}\, \delta(x^-)\, \log(|x^{\perp}|)\, (\ud x^-)^2 \ ,
\end{equation}
where $P_{-}$ denotes the lightfront energy of the shock wave background. 

As before, we want to write down the final state for a particle crossing a shock wave, neglecting radiation; by \eqref{one} this is fixed by the 2-point amplitude on the background. Exploiting a double copy for shock wave space-times, the electromagnetic and gravitational cases can be treated uniformly by writing the 2-point amplitude as~\cite{tHooft,Jackiw:1991ck,Kabat:1992tb,Adamo:2021rfq,Adamo:2021jxz}:
\begin{equation}\label{eq:eik-stato}
    \bra{p'}\mathcal{S}\ket{p}=\frac{p_{+}}{\hbar^{2}} \hat{\delta}\left(p_{+}-p_{+}^{\prime}\right) \int \mathrm{d}^{2} x^{\perp}\, \mathrm{e}^{\mathrm{i}\, q_{\perp} x^{\perp}/\hbar+\mathrm{i}\, \chi(x^{\perp})/\hbar} \, ,
\end{equation}
where $q_{\perp}:=p_{\perp}'-p_{\perp}$ and $\chi(x^{\perp})$ is a theory-dependent eikonal phase whose explicit form for electromagnetism and gravity will be given below. The final semi-classical state for a particle crossing a generic shock wave background is thus
\begin{equation}
    \mathcal{S}\ket{\Psi}=\int \ud\Phi(p,p')\, \phi_{b}(p)\, \frac{p_{+}}{\hbar^{2}}\, \hat{\delta}\!\left(p_{+}'-p_{+}\right) \int \mathrm{d}^{2} x^{\perp}\, \mathrm{e}^{\mathrm{i}\,q_{\perp} x^{\perp}/\hbar+\mathrm{i}\, \chi(x^{\perp})/\hbar}\, \ket{p'} \ .
\end{equation}
We can integrate over $p_{+}$ directly while the $p_{-}$ integral is fixed by the on-shell delta function in the measure. Changing variables from $p_{\perp}$ to $q_{\perp}$ leaves
\begin{equation}
    \mathcal{S}\ket{\Psi}=\frac{1}{\hbar^2}\int \!\ud\Phi(p')\, \e^{\mathrm{i}\, p \cdot b/\hbar} \int\!\mathrm{d}^{2} x^{\perp} \!\!\int \!\mathrm{d}^{2} q_{\perp} \: 
    \phi(p_{+}^{\prime},p'_{\perp}+q_{\perp})\,
    \mathrm{e}^{\mathrm{i}\, q_{\perp} (x-b)^{\perp}/\hbar+\mathrm{i}\, \chi(x^{\perp})/\hbar}\, \ket{p'}  \ .
\end{equation}
Expressed in this form, the final semi-classical state agrees, in the conservative sector and in the massless limit of one particle, with the final state describing two point particles,recently studied from the point of view of on-shell amplitudes in~\cite{Cristofoli:2021jas}. Following the same strategy as employed there, we can evaluate the remaining $x$ and $q$ integrals using a stationary phase approximation.  This is motivated by the fact that in the classical limit the eikonal phase (equal to the radial action to this order) is large compared to $\hbar$.

The stationary phase conditions in the $(x,q)$ plane are given by
\begin{equation}
x_{*}^\perp = b^{\perp} -\frac{\partial}{\partial {q_{\perp}}}\chi(x_{*}) \quad , \quad
q_{* \perp}=-\partial_{\perp}\chi(x_{*})    \ .
\end{equation}
The first equation defines the so called \textit{eikonal impact parameter}, while the second localizes the exchange momentum to the derivative of the phase\footnote{By considering the eikonal phase at LO in the coupling one can safely neglect the derivative with respect to $q_{\perp}$. However, at higher orders this term is non vanishing and is crucial in deriving the correct observables. For further details, we refer the reader to Section 4 of \cite{Cristofoli:2021jas}.}. As a result, we have that the final \emph{semi-classical} state on a shock wave background may be written as
\begin{equation}\label{sevgilim}
    \mathcal{S}|\Psi\rangle=\int \ud \Phi(p^{\prime})\,   \e^{\mathrm{i}\, \delta(x_{*},q_{*})/\hbar}
    \phi(p_{+}^{\prime},p'_{\perp}-\partial_{\perp} \chi(b))\,
    \left|p^{\prime}\right\rangle \ ,
\end{equation}
where $\delta(x_{*},q_{*})$ is the remaining contribution in the exponent from the stationary phase argument; it will not contribute to the observables of interest.

Note that \eqref{sevgilim} is formally equivalent to the corresponding final state \eqref{memo} for a particle crossing an electromagnetic plane wave, if we replace the derivative of $\chi$ with $a(\infty)$. This suggests that the non-vanishing value of the eikonal phase can be interpreted as a velocity memory effect in a shock wave background.
As for the explicit eikonal phases in electromagnetism and gravity, these are related by a double copy~\cite{Cristofoli:2020hnk,Shi:2021qsb} and are given by
\begin{equation}\label{two-chis}
\chi_{\mathrm{EM}}(x^{\perp})=\frac{e^2 \, \log(|x^{\perp}|)}{2\pi}\,, \qquad 
\chi_{\mathrm{GR}}(x^{\perp})=-4G_{N}\, p \cdot P \, \log(|x^{\perp}|) \,.
\end{equation}
The calculation of the classical impulse proceeds by inserting \eqref{sevgilim} into \eqref{general}, and is equivalent to that in a plane wave. (In fact one can read off the result from the argument of the wavepacket in \eqref{sevgilim}.) For the case of an electromagnetic shock wave we obtain, using (\ref{two-chis}),
\begin{equation}\label{final-U1}
    \Delta p^{\mu}=-\frac{e^2\, b^{\mu}}{2\pi\, |b^2| }+\frac{e^4\,n^{\mu}}{8 \pi^2\, p_+\,|b^2|} \,,
\end{equation}
in which we have used $p\cdot b=0$ and defined $b \cdot b:=-|b|^2$. Finally, the impulse in the gravitational case is
\begin{equation}\label{final-GR}
    \Delta p^{\mu}=\frac{4 G_{N}\,p \cdot P \, b^{\mu}}{|b^2|}+\frac{8\, G^2_{N}\, (p \cdot P)^2 \, n^{\mu}}{p_{+}\,|b^2|} \;.
\end{equation}
We find in both cases agreement with the literature \cite{Adamo:2021jxz,Shore:2018kmt}. Incidentally, both (\ref{final-U1}) and (\ref{final-GR}) uniquely determines the geodesics on such backgrounds.



\section{Conclusion}

We have generalized the KMOC framework~\cite{Kosower:2018adc} by providing a systematic way to extract classical observables from the classical limit of scattering amplitudes on curved backgrounds. We considered asymptotically flat gravitational backgrounds which are exact solutions of the Einstein equations admitting an $S$-matrix, and electromagnetic backgrounds which are exact solutions to the Maxwell equations, both sourced and source-free. To relate observables to amplitudes computed on such backgrounds, we followed the same strategy as adopted in~\cite{Cristofoli:2021jas}, which consists of first determining the final state in terms of amplitudes and only afterwards evaluating the mean value of an operator by extracting classical observables.As applications, we considered two non-trivial backgrounds: vacuum plane waves and shock waves, both in electromagnetism and in gravity. In the plane wave case, we saw that $2$-point amplitudes provide classical velocity memory effects, finding agreement with the literature, including the contribution coming from zero frequency modes of the emitted field that has not previously been obtained from KMOC.
In the case of a shockwave background, we traced the form of the final state back to a recent conjecture for the final semiclassical state of two interacting particles \cite{Cristofoli:2021jas} and derived the impulse by reconstructing all the geodesics on such space-times. Interestingly, we have seen that the final semiclassical state assumes a universal form given by a simple shift in the exchanged momentum of a free wavepacket, regardless of its definition on a plane wave or shockwave background, either electromagnetic or gravitational. Based on \cite{Adamo:2021rfq}, we expect the same to hold also for a particle on a linearized Schwarzschild and Kerr background since their respective $2$-points will still be given by (\ref{eq:eik-stato}) but for a different eikonal phase. For plane waves, this is due to the high degree of symmetry present, which renders the conservative physics semiclassical-exact. For Schwarzschild, Kerr and shockwaves, the result can be interpreted as the statement that the final states describe conservative semiclassical dynamics for $2$ particles on a flat space time, \emph{in the probe limit} of one of the two \cite{Cristofoli:2021jas}.

While our focus here was on conservative physics, the general formalism we developed relates classical radiative physics to higher-point scattering amplitudes on curved backgrounds. This is apparent through the general expression \eqref{npoint} for the final state.
A clear next step would be to compute leading radiative effects on plane wave and shockwave backgrounds using 3-point amplitudes; many of the ingredients for this have already been found in both electromagnetism~\cite{Adamo:2021jxz,Fedotov:2022ely} and gravity~\cite{Lodone:2009qe,Adamo:2017nia,Adamo:2020qru}. Beyond 3-points, explicit calculations for amplitudes in the background field formalism become increasingly difficult. However, \emph{all-multiplicity} formulae for scattering amplitudes in chiral plane wave backgrounds have recently been found with twistor methods~\cite{Adamo:2020syc,Adamo:2020yzi,Adamo:2022mev}, showing that there is scope for studying high-order radiation effects with this formalism.

We note that for many years an obstacle to pushing plane wave impulse calculations in QED~\cite{Ilderton:2013dba,Ilderton:2013tb} beyond $\mathcal{O}(e^2)$ was the difficultly in establishing the full quantum result, corresponding to higher point and higher loop amplitudes, before the classical limit was taken. By instead reconsidering such calculations within the framework proposed here, we hope to reduce this complexity. While the ultimate goal is gravitational observables, recent \emph{all}-orders classical and quantum results in QED~\cite{Torgrimsson:2021wcj} can provide a benchmark for this program.

Another generalization would be to investigate other backgrounds, particularly black hole space-times in gravity. While these do not have a well-defined $S$-matrix, scattering `at large distances' still makes sense (cf., \cite{Adamo:2021rfq}). However, wavefunctions on these backgrounds are not WKB-exact, so calculations will be substantially more involved. A broad class of WKB-exact backgrounds is still available in terms of `impulsive pp-waves,' which have a physical interpretation in terms of ultra-boosted beams. It would also be interesting to explore in more detail the relation between this formulation of KMOC on curved backgrounds, and recent works based on the exponentiation of the $S$-matrix defined on a flat background \cite{Cristofoli:2021jas,Damgaard:2021ipf,DiVecchia:2022owy}.

More generally, we hope that this formalism will enable an interaction between QFT on strong backgrounds and the modern `amplitudes program,' much as the original KMOC formalism has leveraged the successes of the amplitudes program into classical physics.

\acknowledgments

We are grateful to Riccardo Gonzo, Donal O'Connell and Matteo Sergola for useful conversations and comments on the draft. The authors are supported by a Royal Society University Research Fellowship (TA) and by the Leverhulme Trust, RPG-2020-386 (TA and AC).

\appendix
\section{Classical motion in plane waves}\label{AppA}

We briefly review how contributions linear and quadratic in the coupling, and related to memory effects in the impulse, can be derived from the classical equations of motion of a particle on a given plane wave, first in electromagnetism and then in gravity. The field strength for a given electromagnetic plane wave is $F_{\mu\nu}(x) = n_{\mu}\, E_\nu(x^{-})-{E}_\mu(x^{-})\, n_{\nu}$, in which $E_\mu = (0,0,E_\perp)$ are the compactly supported electric field components of the wave. A particle of charge $e$ and mass $m$, moving in the wave obeys the Lorentz force equation
\be\label{Lorentz}
    \dot{P}_{\mu} (\tau) = \frac{e}{m}\, F_{\mu\nu}(x(\tau))\,P^\nu(\tau) \;,
    \ee
in which $x^\mu(\tau)$ is the orbit parameterised by the proper time $\tau$, $P_\mu=m {\dot x}_{\mu}$ is the on-shell physical momentum, and a dot denotes a derivative with respect to the proper time. Assuming the plane wave background is fixed, one could study the dynamics of a charged particle by inserting the ansatz $x=x_{(0)} + e x_{(1)} +e^2 x_{(2)}+\ldots$ in the equations of motions and by solving order by order in the coupling, without expanding in the background field. One would find in this approach that the $\mathcal{O}(e^2)$ results are in fact exact. To see this in a simple, and manifestly gauge-invariant, manner we dot $n_\mu$ into (\ref{Lorentz}), finding $n\cdot \ddot{x}=0 \rightarrow x^- = p_+\tau/m$, in which $p_\mu$ is the free momentum in the asymptotic past. One can thus change variables from $\tau$ to lightfront time $x^-$, upon which (\ref{Lorentz}) becomes \emph{an ODE} in $x^-$ which is solved by exponentiation:
\be\label{expres}
    \frac{\d P_\mu}{\ud x^{{-}}}(x^-) = \frac{e}{p_+} F_{\mu\nu}(x^-)P^\nu(x^-) \quad \rightarrow \quad P_\mu(x^-) = \mathcal{T} \exp\bigg[\frac{e}{p_+} \int_{x^-_i}^{x^-}\!\ud s\, F(s)\bigg]_{\mu\nu}p^\nu \;.
\ee
The time-ordered exponential truncates at second order since contractions with more than two field strengths are vanishing. Hence the integrals are easily evaluated and the exact result is
\be
    P_\mu(x^-) = p_\mu - e a_\mu(x^-) + n_\mu \frac{2ea(x^-)\cdot p - e^2a(x^-)\cdot a(x^-)}{2p_+} \;,
\ee
in which $a_\mu$ is the work done on the charged particle as it crosses the plane wave, given by (\ref{a-int-E}). Evaluating the final momenta at $x^{-}=+\infty$, one immediately recovers the expression (\ref{eq:QEDresult}) for the impulse as a function of $a_\mu(\infty)$ which was derived from a $2$ point amplitude in electromagnetism. For further examples of classical particle dynamics in strong electromagnetic backgrounds see~\cite{Gonoskov:2021hwf}.

We now consider the corresponding change in momentum for a particle on the gravitational plane wave background (\ref{altingun}). The classical dynamics is governed by the geodesic equation
\begin{equation}\label{eq:geo}
\dot{P}^{\mu}(\tau)=-\frac{1}{m}\Gamma_{\nu \rho}^{\mu}(x(\tau)) P^{\nu}(\tau) P^{\rho}(\tau) \ .
\end{equation}
On a plane wave, the only nonvanishing Christoffel symbols are
\begin{equation}
    \Gamma_{--}^{a}=-H_{\: b}^{a}\left(x^{-}\right) x^{b}, \quad \Gamma_{--}^{+}=-\frac{1}{2} \dot{H}_{a b}\left(x^{-}\right) x^{a} x^{b}, \quad \Gamma_{-a}^{+}=-H_{a b}\left(x^{-}\right) x^{b} .
\end{equation}
The vanishing of $\Gamma^{-}_{\mu \nu}$ implies ${\dot P}^-=0$ as in the electromagnetic case, so that proper time can again be traded for lightfront time $x^-$. Despite the additional dependence on the transverse coordinates $x^a$ in the geodesic equation (\ref{eq:geo}), the formal exponential solution
\begin{equation}
 P^{\mu}(x^-)= \mathcal{P} \exp \bigg(-\frac{1}{p_+}\int_{-\infty}^{x^-} \mathrm{d} y^- \: \Gamma_{\nu \rho}^{\mu} \dot{x}^{\rho}(y^-)\bigg) p^{\nu} \;,
\end{equation}
still truncates and leads to the \emph{explicit} solution 
\begin{equation}
\begin{split}
P_+(x^{-}) &= p_+ \;, \\
P_{a}(x^{-}) &= p_{i} E_{a}^{i}(x^{-})+p_{+} \sigma_{a b}(x^{-}) x^{b}(x^{-})  \ ,\\
P_{-}(x^{-}) &=\frac{m^{2}}{2 p_{+}}+\gamma^{i j}(x^{-}) \frac{p_{i} p_{j}}{2 p_{+}}+\frac{p_{+}}{2} \dot{\sigma}_{b c}(x^{-}) x^{b}(x^{-}) x^{c}(x^{-})+p_{i} \dot{E}_{b}^{i}(x^{-}) x^{b}(x^{-}) \ ,
\end{split}
\end{equation}
where $\sigma_{a b} \left(x^{-}\right):=\dot{E}_{a}^{i} E_{i b}$ and $\gamma^{ij}(x^{-}):=E^{a(i}E^{j)}_{\:a}$. It is easy to see that $P^2=g^{\mu\nu}P_{\mu} P_{\nu}=0$ for all $x^-$, thanks to the identity $\dot{\sigma}_{ab}=\dot{E}^{i}_{a}\dot{E}_{ib}-H_{ab}$. Using (\ref{grbc}) in the limit $x^{-}\to \infty$ one easily recovers (\ref{no}), in agreement with the impulse derived from a $2$-point amplitude in gravity.

\bibliographystyle{JHEP}

\providecommand{\href}[2]{#2}\begingroup\raggedright\endgroup

\end{document}